\documentclass[manuscript,nonacm]{acmart}
\AtBeginDocument{%
  }

\acmConference[ACM]{Conference on Human Factors in Computing Systemsl}{
  2026}{Barcelona, Spain}
\usepackage{tabularx}
\usepackage{booktabs}
\usepackage{array}
\usepackage{makecell}
\setcopyright{none}
\usepackage{array}
\usepackage{makecell}
\begin{document}

\title{Relationship-Centered Care: Relatedness and Responsible Design for Human Connections in Mental-Health Care}



\author{Shivam Shukla}
\affiliation{%
  \institution{University of California, Santa Cruz}
  \country{USA}
  }
\email{sshukla3@ucsc.edu}

\author{Emily Chen}
\affiliation{%
  \institution{University of California, Santa Cruz}
  \country{USA}
}
\email{eche27@ucsc.edu}

\author{Mahnaz Roshanaei}
\affiliation{%
  \institution{Stanford University}
  \country{USA}
}
\email{mroshana@stanford.edu}

\author{Magy Seif El-Nasr}
\affiliation{%
  \institution{University of California, Santa Cruz}
  \country{USA}
}
\email{mseifeln@ucsc.edu}






\renewcommand{\shortauthors}{Shivam et al.}

\begin{abstract}
There has been a growing research interest in Digital Therapeutic Alliance (DTA) as the field of AI-powered conversational agents are being deployed in mental health care particularly those delivering CBT (Cognitive Behaviour Therapy). Our proposition argues that the current design paradigm which seeks to optimize the bond between a patient in need of support and an AI agent, contains a subtle but consequential trap: it risks producing an “appearance of connection” that unintentionally disrupts the fundamental human need for relatedness, which potentially displaces the authentic human relationships upon which long-term psychological recovery and overall well-being depends. We propose a reorientation from designing artificial intelligence tools that simulates relationships to designing AI that scaffolds them. To operationalize our argument, we propose an interdisciplinary model that translates the Responsible AI Six Sphere Framework through the lens of Self-Determination Theory (SDT), with a specific focus on the basic psychological need for relatedness. The resulting model offers the technical and other clinical communities a set of relationship-centered design guidelines and relevant provocations for building AI systems that function not just as companions, but as a catalyst for strengthening a patient's entire relational ecology; their connections with therapists, caregivers, family, and peers. In doing so, we discuss a model towards a more sustainable ecosystem of relationship-centered AI in mental-health care. 

\end{abstract}

\ccsdesc[500]{Human-centered computing~Collaborative and social computing}
\keywords{Digital Therapeutic Alliance, Self-Determination Theory, Relational Ecology, Mental Health, Responsible AI, Human-AI Interaction, Relational Displacement}

\begin{teaserfigure}
  \includegraphics[width=\textwidth]{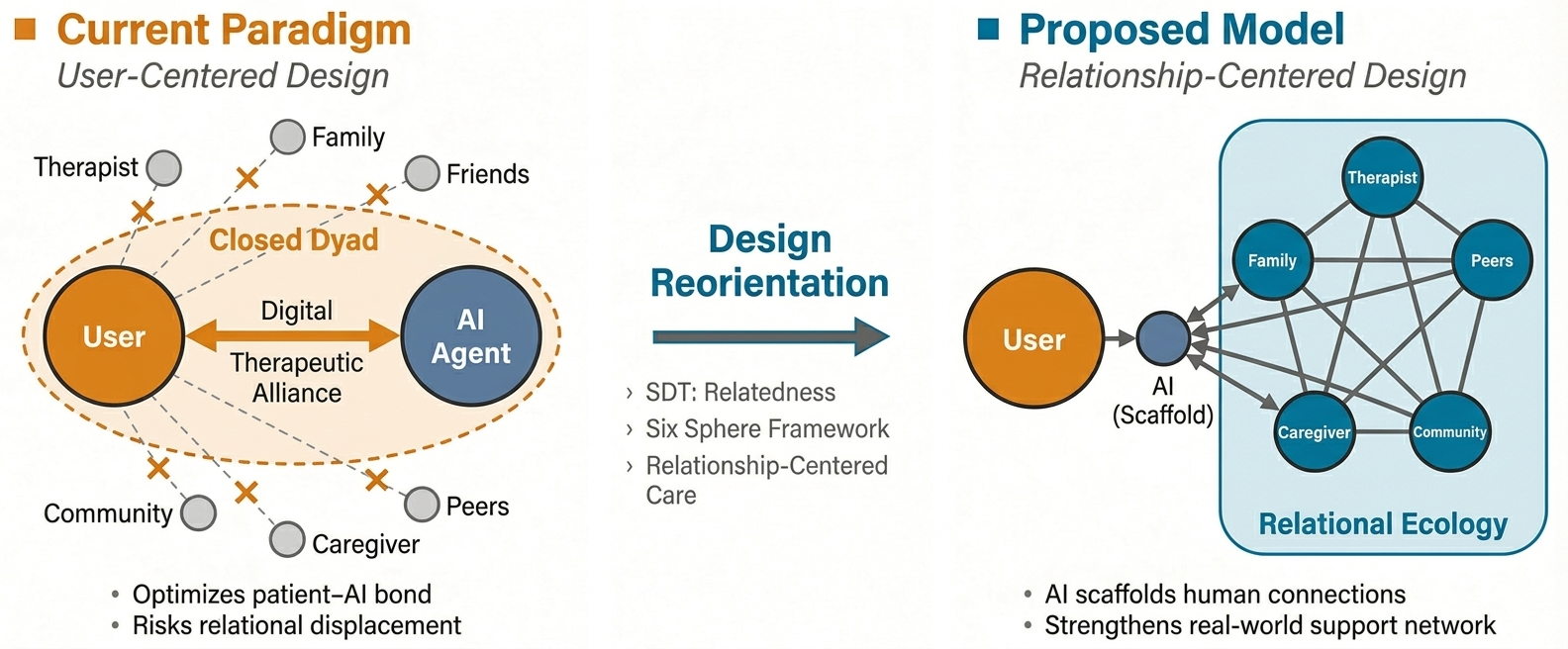}
  \caption{From simulating relatedness to scaffolding them}
  \Description{Enjoying the baseball game from the third-base
  seats. Ichiro Suzuki preparing to bat.}
  \label{fig:teaser}
\end{teaserfigure}


\maketitle

\section{The Emotional  Design Trap in Simulating Relatedness}
The rise of AI-based systems for mental-health shifts the landscape of therapeutic care. Platforms like Limbic, Woebot, and Wysa have demonstrated considerable promise in delivering evidence-based interventions, particularly Cognitive Behavioral Therapy, showing efficacy in reducing symptoms of depression and anxiety at scale \citep{fitzpatrick2017delivering, inkster2018empathy, mcfadyen2024ai}. As this field is evolving a central focus has been the cultivation of Digital Therapeutic Alliance (DTA), with studies suggesting that users can form timely bonds with an AI conversational agent that is quantitatively (therapeutic alliance scores) comparable to the alliance formed with a human therapist \cite{beatty2022evaluating}.  Content analysis of user transcripts with agents reveals unprompted elements of bonding, including expressions of gratitude, self-disclosed impact, and even the personification of an AI agent \citep{beatty2022evaluating}. In a recent integrative review done by Malounie et al. on 28 studies in the DTA in AI-driven psychotherapeutic interventions, they have found that users are prone to make a bond with conversational agents with empathic tones and anthropomorphic features \citep{malouin2025does}. Johnson et al. also found users returning back to share updates with chatbots about their emotions and experiences, reflecting about their long term engagement and sense of relatedness with the system \citep{ta2022assessing}. This bond is a crucial driver of engagement, adherence, and clinical outcomes reflecting the critical role of alliance in traditional psychotherapy \cite{malouin2025does}. And now since the focus on the human-AI relation has been increased it has also raised ethical and relational questions.  A growing body of researchers and practitioners has raised concerns about the current trajectory. For example, through systematic review of embodied AI tools in the field of Psychiatry, Psychology and Psychotherapy Fiske et. al have warned about the ethical implications cautioning that the relationship between patient and provider should not be eroded \citep{fiske2019your}. Brown and Halpern argued that chatbots cannot replace the critically therapeutic aspects  of human interaction in the pursuit of more inclusive mental-health care \cite{brown2021ai}. Tavory has also proposed that the dominant “responsible AI” approach which governs most regulatory and ethical guidance is still insufficient because AI has tendency of emotional manipulation and falls short because the impact of Artificial Intelligence on human-relationships is being overlooked \cite{tavory2024regulating}.  


We contend that this pursuit contains a subtle but critical design trap. By optimizing for the feeling of connection with a non-human agent, we risk the experience of simulated empathy with the genuine satisfaction of the basic psychological need of relatedness. Self-Determination Theory, one of the most robustly evidence-based theory of human-motivation and well being \citep{ryan2000self}, defines relatedness as the fundamental psychological need to feel connected with others, which means to experience a sense of belonging within social context. Decades of research have linked the satisfaction of this need to enhanced self-motivation, psychological health, and even greater longevity; considered more powerful than diet and exercise \citep{peters2018designing}. Critically, not all social interactions satisfy this need evidenced by the work done by Peter et al. on METUX model \citep{peters2018designing}. They propose “it is essential to design for genuine relatedness, rather than the mere semblance of connection \citep{peters2018designing}. " Even a well-designed, AI system can only ever offer the appearance of relatedness. The risk, therefore, lies in what we term as a \textbf{relational displacement}. When a patient, particularly one in a vulnerable psychological state, finds one click accessible, non-judgmental, and empathetic AI companion, they may begin to substitute interactions with the AI for the more complex and often challenging work of maintaining authentic human relationships. This is not a failure of the patient, but a predictable outcome of a system designed for maximum engagement within the patient-AI dyad. Peters et al. explicitly warned that over-engagement with even a highly need-satisfying application can lead to a state where “important activities get crowded out leading to drops in relatedness as human relationships are being ignored \citep{peters2018designing}.” In the context of clinical mental health, if the artificial intelligence becomes the primary repository for a patient’s emotional expression and processing, it may inadvertently weaken their connections to the very human support systems which are therapists, family members, peers; that are essential for psychological well-being, recovery and real-world resilience.  


\section{Bridging the Frameworks: From SDT to the Six Spheres}

In order to move beyond critique, we move toward constructive design guidance, we propose a model that reorients the goal and position of AI in mental-healthcare. As we discussed above, this requires a shift  in our unit of analysis from the patient-AI dyad to the patient’s entire relational ecology. The Six Sphere Model provides a structured approach to evaluating the ethical and well-being impacts of a technology across six concentric spheres of experience: Adoption, Interface, Task, Behaviour, Life, and Society \citep{peters2020responsible}. Crucially, this model is a direct extension of the author’s earlier work on the METUX (Motivation, Engagement, and Thriving in User Experience) model, which was co-authored by the co-developer of SDT, Richard M. Ryan \citep{peters2018designing, ryan2000self}. The METUX model was explicitly designed to operationalize SDT’s  three basic psychological needs (autonomy, competence, and relatedness) within these distinct spheres of technology experience. The Six Sphere framework, therefore, is already implicitly grounded in SDT’s model and could be translated and utilized for our proposition.  


By centering relatedness, we align the framework directly with the principles of Relationship-centered care (RCC) and the workshop's core theme of designing for human connection in healthcare \citep{beach2006relationship}. Furthermore, the Six Sphere framework’s original case study already hints at the relational concern we are discussing here. Under the principle of Beneficence, Peters et al. note that even sophisticated AI systems “lack human empathy and are at best able to mimic these traits \citep{peters2020responsible}." They raise therapist’s concern about the loss of nonverbal cues, the absence of physical human interaction, and the resulting feelings of “alienation and devaluation.” Our proposition takes these scattered observations and systematizes them into a human-centered design model. 

\section{A Relational Ecology Model: Centering Human Connection}

As detailed in Table 1, we present a relational ecology model that offers design guidelines and provocations intended to inspire new thinking within the HCI and clinical communities. From the moment of \textbf{Adoption}, the system should be framed as a bridge to human care, setting an honest relational expectation before the first interaction begins. At the \textbf{Interface level}, this means replacing inward-facing prompts to share feelings with the AI with outward-facing affordances. At the \textbf{Task level}, standard CBT exercises need not remain private transactions with an algorithm. Moving into the \textbf{Behaviour} sphere, the primary outcome the system is optimized to produce should not be user adherence to an AI-delivered program but the quality and frequency of the user's supportive human interactions, with the AI actively prompting engagement with therapists, family members, and community rather than treating such engagement as incidental to its own metrics. At the \textbf{Life} level, symptom reduction measured through standardized scales remains a legitimate clinical goal, but it is an insufficient one if achieved through a dynamic that silently rewards isolation within the user-AI dyad. And at the \textbf{Society level}, the stakes widen to their fullest extent: if AI-mediated care becomes the dominant mode of therapeutic support at population scale without being deliberately designed to scaffold human connection.

\begin{table*}[t]
\small
\renewcommand{\arraystretch}{1.3}

\begin{tabularx}{\textwidth}{
>{\raggedright\arraybackslash}p{2.2cm}
>{\raggedright\arraybackslash}X
>{\raggedright\arraybackslash}X
>{\raggedright\arraybackslash}X
}

\toprule
\textbf{Sphere} &
\textbf{Conventional Goal (User-Centered)} &
\textbf{Proposition (Relationship-Centered)} &
\textbf{Design Provocation (Example)} \\
\midrule

\textbf{Adoption} &
Persuade patient about the effectiveness and empathic capacity of AI. &
Along with effectiveness and empathic capacity, present the AI as a tool to enhance connections with human supporters. &
How might marketing and onboarding materials explicitly position the AI as a bridge to human care, setting authentic expectations that scaffold? \\

\textbf{Interface} &
Design for seamless, empathetic conversation with the AI agent. &
Design interface features that facilitate communication with other people, not just AI. &
Instead of one-way HAI interaction prompts like ``share your feelings,'' what about a ``draft a message to someone you trust'' feature? \\

\textbf{Task} &
Deliver AI-led CBT exercises (e.g., automated thought records, mood tracking). &
Design AI-guided tasks that also involve and orient toward another person (e.g., preparing for a difficult conversation, practicing interpersonal gratitude, rehearsing assertiveness skills). &
Can a standard CBT thought record be reframed or practiced as a collaborative exercise with a trusted human (of the patient’s choice), where the AI mediates rather than replaces the therapeutic dialogue? \\

\textbf{Behaviour} &
Increase patient adherence to the AI-delivered therapeutic program. &
Increase the quality and frequency of the patient’s supportive human interactions as the primary behavioral outcome. &
How might the AI prompt patients to schedule check-ins with their therapist, reach out to a supportive family member, or engage in a community activity? \\

\textbf{Life} &
Reduce the patient’s symptoms of depression and anxiety as measured by standardized scales. &
Enhance the patient's sense of belonging and reduce these symptoms by strengthening their real-world social support network. &
Does the system track and encourage activities that build social capital and authentic connection, or does it silently/inadvertently reward isolation with the AI? \\

\textbf{Society} &
Increase access to mental healthcare at population scale. &
Along with the increase in accessibility to MH care, reduce the growing problem of people feeling disconnected and lonely. &
What is our collective responsibility to prevent a future where therapeutic care is primarily mediated by non-human agents, and what design safeguards can we build today? \\

\bottomrule
\end{tabularx}

\caption{Shifting from user-centered to relationship-centered design perspectives in AI-mediated mental health systems.}
\label{tab:relationship-centered}
\end{table*}

\section{Discussion and Conclusion}
This submission has argued for a fundamental reorientation in how the technical community approaches the design of AI for clinical mental health interventions. By moving from a user-centered model that prioritizes the human-AI Interaction to a relationship-centered model that prioritizes the vitality of patient's entire relational ecology, we can begin mitigate the ethical risks of relational displacement and design for more sustainable and holistic approach towards well-being. The model we presented here offers a practical and grounded approach for achieving this goal. The implications of this model for the community are significant and multifaceted. First, it challenges us to develop \textbf{new metrics for success.} Instead of optimizing for session length, daily active patients, or engagement scores with the AI system, we should be developing and validating measures of the system's impact on the quality and quantity of patient's real world social interactions. Second, it calls for \textbf{new design patterns} that actively orient patients outward toward human support networks. Third, it demands a more \textbf{interdisciplinary design process, } one that brings together not only engineers and clinicians, but also ethicists, social workers, people  with lived mental health challenges, in the spirit of participatory design and design justice that the Six Sphere framework also advocates \citep{spinuzzi2005methodology, costanza2020design, peters2020responsible}.

We aim to test this model as our future work and we also acknowledge that this position has its own limitations. For individuals in acute crisis, or those with no access to human support, an AI companion may be the only available resource, and designing it to deviate emotional engagement could be harmful and not efficient utilization of the technology itself. Our model is not a prescription for all contexts, but a provocation for the existing design paradigm. The goal is not to eliminate the Digital Therapeutic Alliance, but to elevate it to the higher-level objective of fostering human-to-human connections. 

As we stand at the beginning of new era in mental healthcare, we have a critical opportunity and profound responsibility to shape the role that AI will play.  Because at the end of the day we are creating these technologies for human flourishing and human connections are inherently important; \textit{"When the computer dies, another computer does not cry."}

\bibliographystyle{ACM-Reference-Format}
\bibliography{Human-AI-Relational-Ecology}

\end{document}